\documentclass{aa}
\usepackage{graphicx,amssymb}

\begin{document}

%\thesaurus{05(08.02.2; 08.02.4; 08.02.7; 08.09.2; 10.15.2)}

\title{The blue straggler S\,1082: a triple system in the old open 
       cluster M\,67\thanks{The William Herschel Telescope, the Isaac Newton
       Telescope and the Jacobus Kapteyn Telescope are operated on the island
       of La Palma by the Isaac Newton Group in the Spanish Observatorio del
       Roque de los Muchachos of the Instituto de Astrofisica de
       Canarias. The Dutch 0.91 Telescope is operated at La Silla by the 
       European Southern Observatory.}}

\titlerunning{The blue straggler S\,1082: a triple system in the old
       open cluster M\,67}

\author{Maureen van den Berg\inst{1} \and Jerome Orosz\inst{1} 
\and Frank Verbunt\inst{1} \and Keivan Stassun\inst{2}}

\authorrunning{Maureen van den Berg et al.}

\offprints{Maureen van den Berg}

\mail{m.c.vandenberg@astro.uu.nl}

\institute{   Astronomical Institute, Utrecht University,
              P.O.Box 80000, 3508 TA Utrecht, The Netherlands
         \and Department of Astronomy, University of Wisconsin-Madison,
              475 N Charter St, Madison WI~53706, USA
              }

\date{Received date \today / Accepted date}

\abstract{We present a photometric and spectroscopic study of the blue
straggler S\,1082 in the open cluster M\,67.  Our observations confirm
the previously reported 1.07 day eclipse light curve and the absence
of large radial-velocity variations of the narrow-lined star. However,
we find two more spectral components which {\em do} vary on the 1.07
day period. We conclude that the system is triple. We fit the light
and radial-velocity curves and find that the total mass of the inner
binary is more than twice the turnoff mass and that the outer
companion to the binary is a blue straggler on its own account. We
briefly discuss formation scenarios for this multiple system.
 }

\maketitle

\section{Introduction}

Blue stragglers are stars that are bluer and more luminous than the
main-sequence turnoff of the population to which they belong. Hence
they appear to be either younger than the other stars or have
main-sequence lifetimes that exceed those of stars of similar
mass. Blue stragglers are easiest to distinguish in globular and open
clusters but are also found among field halo stars.  Formation
scenarios that require anomalous evolution of a single star are
considered less likely than explanations that involve the interaction
of stars in a binary or in a multiple-star encounter.  For example, a
blue straggler may be formed when one star in a binary accretes a
substantial amount of mass from its companion or when two stars merge
either in a binary or in a collision.  In the last case, blue
stragglers carry information about the high-stellar density
environments in which they are found. Reviews on blue stragglers are
given by e.g. Stryker (1993) and Bailyn (1995).  \nocite{stry}
\nocite{bail}

The blue stragglers in M\,67 have received frequent attention. Ten of
them were monitored for radial-velocity variations for nearly twenty
years; this revealed one short-period (4.2 day) eccentric binary and
five circular and eccentric binaries with periods between 850 and 5000
days (Milone \&\ Latham 1992, Latham \&\ Milone 1996). Leonard (1996)
and Hurley et al. (2001) concluded that not one of the proposed
blue-straggler formation mechanisms alone can explain the properties
of these binaries. \nocite{milolath92} \nocite{lathmilo}

S\,1082 is one of the remaining four blue stragglers without an
orbital solution.  Its spectrum shows the lines of an early F-type
subgiant.  These lines show only small radial-velocity variations
($v_{\rm rad}<$7 km s$^{-1}$ peak-to-peak, Mathieu et al.\ 1986)
and little rotational broadening ($v_{\rm rot}\sin i=$4--11 km
s$^{-1}$, van den Berg et al.\ 1999).  This appears to be in
contradiction to the eclipse light curve with a period of 1.07 days
found by Goranskij et al.\ (1992).  The relative velocity $v$ of two
stars in a 1.07\,day binary is $v$=208 ($M/M$$_{\odot})^{1/3}$\,km
s$^{-1}$ where $M$ is the total mass of the binary; the stellar
rotation in such a short-period binary is expected to be synchronised
with the orbit, giving $v_{{\rm rot}}=(R/a)v$, with $a$ the distance
between the binary stars and $R$ the stellar radius.  The eclipse
light curve excludes the low inclination as well as the extreme mass
ratio required to bring the observed velocity limits in agreement with
the predicted values {\em for an F star in the binary}.  We therefore
tentatively conclude that S\,1082 is a (visual or physical)
triple. \nocite{mathea86}

Evidence for a short orbital period is also furnished by the variation
detected in a broad, shallow component (Mathys 1991) in e.g. the
H$\alpha$ line on time scales of hours (van den Berg et al.\ 1999).
Evidence for a wide orbital period of about 1000 days was found from
the radial-velocity variations of the narrow lines of the F-star
(Milone 1991). \nocite{milo}

\nocite{giraea} \nocite{montea}
\nocite{goraea} \nocite{vdbergea} \nocite{math} \nocite{leon}
\nocite{milo} \nocite{hurlea}

We have collected multiband photometry of S\,1082 together with
high-resolution spectra that sample the light curves at different
phases. Our goal was to investigate the eclipses found by Goranskij et
al., to monitor the variability of the second spectral component as
function of photometric phase and to model the light curve together
with the radial-velocity curves.  Sect.~\ref{obs} describes the
observations and data reduction. The results of the spectroscopic and
light curve analysis are presented in Sect.~\ref{results} and
Sect.~\ref{lightc}. In Sect.~\ref{disc} we summarise our
interpretation of the nature of this blue straggler.

\section{Observations and data reduction} \label{obs}

\subsection{Photometry}

S\,1082 was monitored in the $U$, $B$, $V$, $I$ and Gunn $i$ bands on three
occasions (see Table \ref{photlog}). In run 1 we observed the star
during twelve nights with the 0.91m ESO-Dutch telescope at La Silla.
The observing schedule was divided in four blocks of three nights;
every first night the star was observed for an average of 5
consecutive hours in a $U\,B\,V$\,Gunn\,$i$-exposure sequence while every
second and third night typically one to three exposures were taken in
each filter.  During both nights of run 2, S\,1082 was observed for
several hours with the 1m Jacobus Kapteyn Telescope on La Palma.  In
run 3, also on the Jacobus Kapteyn Telescope, we aimed to complete
the phase coverage of the light curve between phases 0--0.3.

\begin{table}
\caption{Log of the photometric observations. For each run we give the
date of the observations, the telescope and filters used and the
typical exposure time for each filter.}
\begin{tabular}{l@{\hspace{0.17cm}}l@{\hspace{0.17cm}}l@{\hspace{0.2cm}}l@{\hspace{0.35cm}}l}
Run & Dates & Telescope  & Filters & t$_{\rm exp}$ (s) \\
    &       &            &                 &                        \\
1   & Feb 8-19 1999      & 0.91m ESO       & $U\,B$     & 300\,120  \\
    &                    & \,Dutch         & $V$\,Gunn $i$  & \,120\,120 \\
2   & Dec 25, 26 1999    & 1m ING JKT      & $B\,V$       & 75\,30     \\
3   & Feb 13-16, 20 2000 & 1m ING JKT      & $U\,B\,V\,I$   & 350\,30   \\
    &                    &                 &           & \,15\,8
\end{tabular}
\label{photlog}
\end{table}

Standard reduction steps of bias subtraction and flatfielding were
performed with IRAF\footnote{IRAF is distributed by the National
Optical Astronomy Observatories, which are operated by the Association
of Universities for Research in Astronomy, Inc., under cooperative
agreement with the National Science Foundation} routines.  Aperture
photometry for all the stars in the field was done with the {\sc
daophot.phot} task.  Differential light curves for each individual
dataset were computed with ensemble photometry (Honeycutt 1992). The
variability properties of the other stars in the fields are discussed
in van den Berg et al. (2001) and Stassun et al. (2001, in
preparation). We refer the readers to these papers for a full
description of the observations and the photometry reduction.

\nocite{honn} \nocite{vdbergea2001bb} \nocite{scar} 

\subsection{Spectroscopy}

\subsubsection{High-resolution spectra} \label{highres_obs}

High-resolution (R$\approx$49\,000) echelle spectra were taken with the
Utrecht Echelle Spectrograph on the 4.2m William Herschel Telescope on
La Palma. S\,1082 was observed on four nights in 1996 and 2000 (see
Table~\ref{speclog} for a log of all spectroscopic observations).

The 1996-spectra were centred on a blue (4250 \AA) and red (5930 \AA)
wavelength.  The 31 lines mm$^{-1}$ grating was used in combination
with the 1024x1024 pixels$^2$ TEK-CCD.  For a full description of the
spectra of run 1 we refer to van den Berg et al.  (1999).

The spectra of 2000 were all taken with the same instrumental setup:
the 79 lines mm$^{-1}$ grating was used with the 2148x2148 pixels$^2$
SITe1-CCD while the spectra were centred on 5584 \AA. In run 2 the
seeing was about 2\arcsec, while light clouds were present during the
start of the run. The slit width was set to 1\arcsec. To secure
stability, no changes were made to the instrumental setup during the
night. In the first two observations of run 5 the seeing was about
2\arcsec; this deteriorated to 3\arcsec\, with cloudiness during the
last two observations. As the slit width was kept fixed at 1\arcsec,
these spectra are of bad quality. In run 6 the seeing was 2\arcsec\
during the first two observations but later improved to 1\farcs6. The
slit width was accordingly changed from 2 to 1\farcs2. Due to the
wider slit these spectra have a lower resolution than the spectra of
run 1 and 5. All frames were exposed for 1200 s, except for those of
run 5 that were exposed for 1800 s to account for the bad seeing
conditions.  During each run we observed radial-velocity
standards. Flatfield images were made with exposures of a Tungsten
lamp. Thorium-Argon lamp emission-line spectra were taken for
wavelength calibration.

Data-reduction was done in IRAF with {\sc ccdred} and {\sc echelle}
routines.  After correcting the frames for the electronic bias and after
flat fielding, spectra for each echelle-order were extracted with
optimal extraction. Towards the red, gaps occur in the wavelength
coverage. For the wavelength calibration, fifth-order polynomials were
fitted in both directions to the positions of the arclines on the CCD;
the maximum residuals to the fit were $\sim$0.0125 \AA\, corresponding
to 0.75 km s$^{-1}$.  Low-order polynomials were fitted to the spectra
for continuum normalization.

\begin{table}
\caption{Log of the spectroscopic observations. For each run we give
the run number, the date of the observations, the wavelength coverage
in \AA, the number of observations, the exposure time in seconds, the
spectrograph used and (if applicable) the radial-velocity standards.}
\begin{tabular}{l@{\hspace{0.1cm}}l@{\hspace{0.18cm}}l@{\hspace{0.23cm}}r@{\hspace{0.19cm}}rll}
 Run  & Dates             & \multicolumn{1}{c}{$\Delta \lambda$} & \#
      & t$_{\rm exp}$ & Inst & RV \\
 &              &
      &     &  &   &         \\
 1    & Feb 28 1996   &  3920-4920
      &  2  &  600, & ues & HD\,136202$^{a}$ \\
 &                &
      &      & \,300 & & \\
 &                &  4890-7940     &  3  &
      360     & ues & HD\,136202 \\
% &                &  3831-5404      &  1  & 60      & isis & -          \\
% &                &      5619-7135     &  1  & 60      & isis & -          \\
 2    & Feb
      16 2000   &  4380-8650     & 17  & 1200    & ues & HD\,89449$^{b}$  \\
      3    & Feb 20 2000   &  3535-5035     &  1  &  900    & ids & -
      \\ 4    & Feb 22 2000   &  3535-5035     &  1  &  900    & ids &
      -          \\  5    & Mar 13 2000   &  4380-8650     &  4  &
      1800    & ues & HD\,89449  \\ 6    & Mar 20 2000   &  4380-8650
      &  5  & 1200    & ues & HD\,89449
\end{tabular}

$^{a}$F8III-IV, $^{b}$F6IV (Simbad)
\label{speclog}
\end{table}

\subsubsection{Intermediate-resolution spectra}

Two intermediate-resolution (R$\approx$3600) spectra were obtained
with the Intermediate Dispersion Spectrograph (IDS) mounted on the
2.5m Isaac Newton Telescope on La Palma, on February 20.90214, 2000
(UT) and February 22.85523 (UT).  The R1200B grating and the EEV10 CCD
combination gave a spectral resolution of $1.19$~\AA\ (FWHM) and a
useful wavelength range of 3533-4825\AA.  The exposure times were 900
seconds each, and the signal-to-noise level in the extracted spectra
ranged from about 65 per pixel near the Balmer jump to about 150 near
4750~\AA.

%A blue and red low-resolution (R$\approx$1000) spectrum was taken in
%run 1 with the ISIS dual-beam spectrograph on the William Herschel
%Telescope. A description of the ISIS spectra can be found in van den
%Berg et al. (1999).

\section{Results} \label{results}

\subsection{Eclipse light curve} \label{eclips}
\begin{figure}
\resizebox{\hsize}{!}{\includegraphics{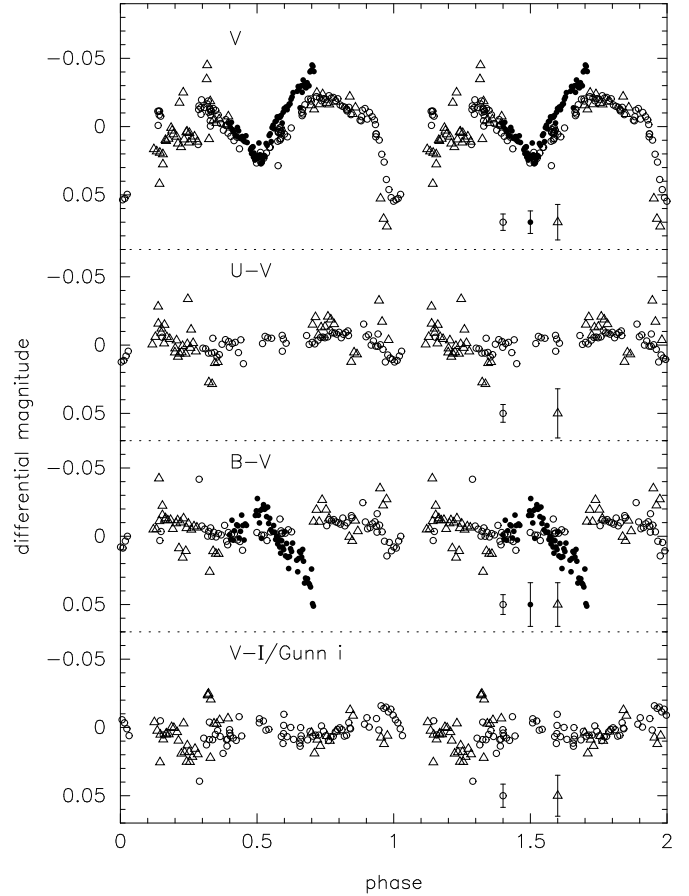}}
\caption{Light and colour curves for S\,1082 folded on the ephemeris
of Goranskij et al.\ (1992). Data from the three runs are indicated
with different symbols: open circles for run 1, filled circles for run
2 and triangles for run 3. A typical error bar is shown for each
run. Magnitude and colours are plotted relative to the average $V$
(11.25), $U$--$V$ (0.45), $B$--$V$ (0.41) and $V$--$I$ (0.57) values in our
measurements.
\label{lc}}
\end{figure}

The new light curves of S\,1082 confirm the findings of Goranskij et
al.\ (1992) and fill in the gap in their data between photometric
phase 0.53 and 0.74 (with phase 0 corresponding to primary
minimum). In Fig.~\ref{lc} our data are folded on their ephemeris for
the primary minimum:
\begin{equation} \label{ephem}
{\rm Min I} = 2\,444\,643.253(5) + 1\fd0677978(50)~E
\end{equation}
Two unequal eclipses with an amplitude of about 0.09 and 0.05 mag in
$V$ are clearly visible with the deeper, primary eclipses occuring
around phase 0 as expected from Eq.~\ref{ephem}.  Goranskij et
al. note that the light curve near second quadrature (phase 0.75) is
systematically higher than near first quadrature (phase 0.25) by 0.01
to 0.02 mag in $V$. This might also be the case in our data, but the
scatter between phases 0.1 and 0.3 makes this difficult to see. The
scatter is probably related to the bad observing conditions (nearby
moon) of the last run.

\subsection{Spectral line profiles} \label{spectra}
\begin{figure}
\resizebox{\hsize}{!}{\includegraphics[angle=-90]{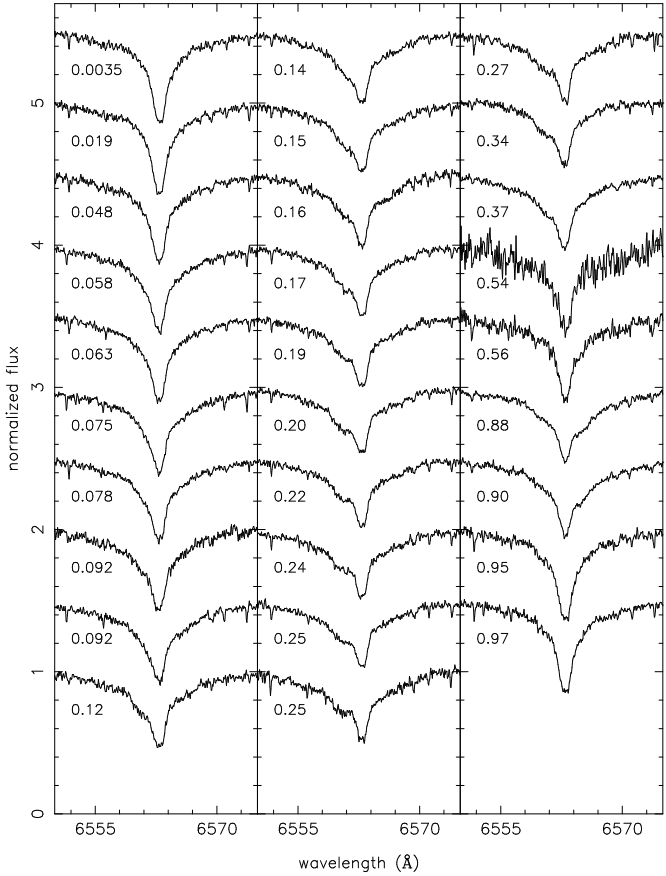}}
\caption{Series of H$\alpha$ profiles in S\,1082. The phase of the
light curve is indicated for each spectrum. Note the central depth
of the line and the asymmetric wings. The observations during phase
0.54 and 0.58 are noisy due to the bad weather conditions of run 5.
The continuum separation between the spectra is 0.5 unit.}
\label{halpha}
\end{figure}

\begin{figure}
\resizebox{\hsize}{!}{\includegraphics[angle=-90]{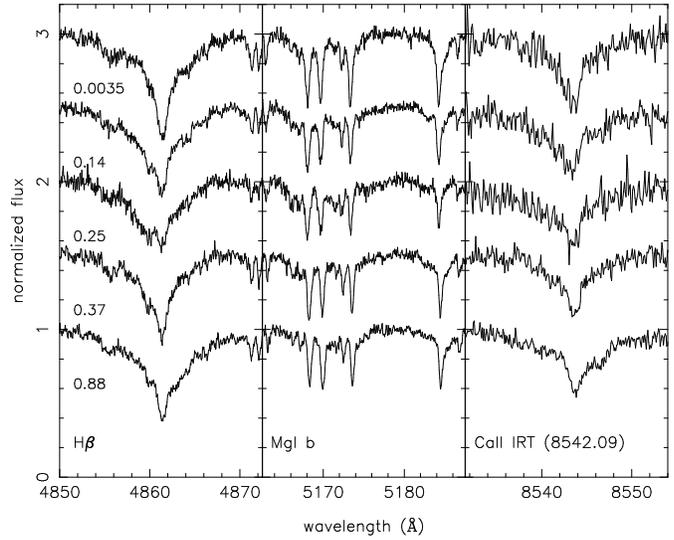}}
\caption{H$\beta$, \ion{Mg}{I}b and \ion{Ca}{II} IRT lines in S\,1082
for a selected set of observations. Light curve phases are
indicated. The continuum separation between the spectra is 0.5 unit.}
\label{otherlines}
\end{figure}

The spectra of S\,1082 are dominated by a narrow-lined component
resembling a star of type early-F.  In addition, the presence of a
broad and shallow component in the high-resolution spectra is
prominent in the H$\alpha$ line (see also van den Berg et al.\ 1999).
When the spectra are arranged in order of light curve phase using
Eq.~\ref{ephem}, it is clear that the position and strength of the
asymmetric absorption in the wings vary regularly (Fig.~\ref{halpha});
the spectra that were included in Fig.~8 of van den Berg et al. (1999)
correspond to photometric phases 0.092, 0.16 and 0.25.  Around
first quadrature (phase 0.25) the depth of H$\alpha$ is smallest while
the asymmetric absorption in the wings is maximally blue-shifted; for
phases between 0.5 and 1 the phase-coverage is sparser, but in the
spectra of phases 0.88 and 0.90 it can be seen that the asymmetric
absorption has moved to the red wing. The timing of the velocity shift
with respect to the eclipses associates the broad-lined feature with
the brighter star in the eclipsing binary (the brighter star must
approach us after the primary eclipse). Similar variable line profiles
are visible in the H$\beta$, \ion{Na}{I} D, \ion{Mg}{I} b and
\ion{Ca}{II} infrared lines (Fig.~\ref{otherlines}).

\begin{figure}
\centering
%\resizebox{\hsize}{!}{\includegraphics{plot_1082_balmer_new4.ps}}
\includegraphics[width=6.5cm]{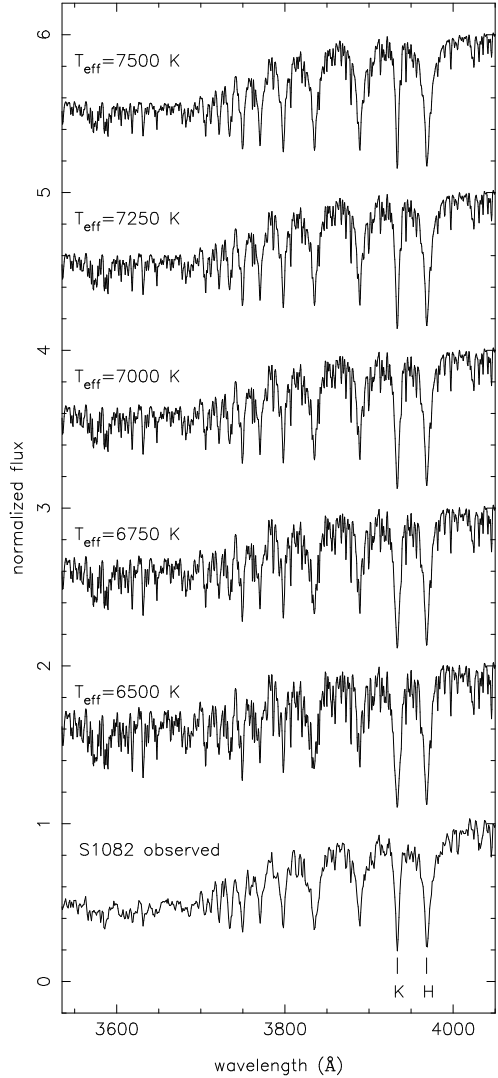}
\caption{Spectrum of S\,1082 compared with synthetic spectra of
various effective temperatures; surface gravity and projected
rotational velocity are fixed at $\log g$=4.5 and $v_{\rm rot}\sin i$=10 km
s$^{-1}$. The spectra are normalised to the flux at 4050 \AA; each
spectrum is offset with 1 unit from the previous spectrum. The
relative strength of the \ion{Ca}{II} K line and the \ion{Ca}{II}
H+H$\epsilon$ blend (marked with 'K' and 'H'), and the contrast in the
Balmer jump (flux near 4050 \AA\ relative to flux near 3600 \AA) are
indicators of temperature.}
\label{lowres}
\end{figure}

The temperature of the narrow-lined star in S\,1082 can be measured
with the low-resolution spectra. We assume that its light is least
contaminated by its companion star(s) in the blue and therefore
concentrate on the region around the Balmer jump. In Fig.~\ref{lowres}
synthetic spectra for various effective temperatures $T_{\rm eff}$ are
compared with the IDS spectrum between 3535 \AA\ and 4050 \AA. The
spectra were computed with model atmospheres for solar metallicity
(Kurucz 1979). The observed spectrum was corrected for the reddening
towards M\,67 ($E(B-V)$=0.032$\pm$0.006, Nissen et al. 1987) and for a
radial-velocity difference with respect to the model spectra. From the
relative strength of the \ion{Ca}{II} H\&K lines, a sensitive
temperature indicator in this region (e.g.\ Gray \&\ Garrison 1989),
and the contrast in the Balmer jump it is clear that the observed
spectrum is hotter than that of a 6500 K star.
%compatible with the F5
%classification of Allen \& Strom (1995) and the temperature estimates
%by Deng et al. (1999), but must be cooler than 7500 K.
\nocite{graygarr} \nocite{nissea} \nocite{kuru}
%\nocite{dengea}

This part of the spectrum was fitted to model spectra with surface
gravity $\log g$ ranging from 0.5 to 5.0 in steps of 0.5, $T_{\rm
eff}$ from 6000 to 8000 K in steps of 250 K and a fixed projected
rotational velocity $v_{\rm rot}\sin i$ of 10 km s$^{-1}$. A free
parameter is a wavelength-independent scale factor ranging from 0.025
to 1 in increments of 0.025 that is a measure of the relative
contribution of the narrow-lined star.  Observed and model spectra are
normalised to the flux at 4050 \AA. A straight line is fitted to the
difference between the observed and each scaled model spectrum. The
model that produces the smallest residuals to the fit has $T_{\rm
eff}$=7500 K, $\log g$=4.5 and scale factor 0.85.  If the temperature
of the hot star is indeed 7500 K, it is a late A rather than an F
star; the value of $\log g$ is close to that of a main-sequence star
for which $\log g \approx 4.25$ (Gray 1992). This shows that the
narrow-lined component dominates the spectrum in the blue and it
implies that one of the individual components of S\,1082 remains a
blue straggler. A more accurate decomposition of S\,1082 is given in
Sect.~\ref{lightc}.  \nocite{montea}

\subsection{Radial-velocity curves}  
\label{highres_ana}

Radial velocities were measured via cross correlation of the
high-resolution spectra with appropriate template spectra
(Table~\ref{vrad}). To determine the radial velocities of the
narrow-lined star in S\,1082 we adopted as templates spectra of F-type
radial-velocity standards observed during the same run (Table
\ref{speclog}). Velocities were derived for each order
separately. Only orders without strong telluric lines were selected,
and included 4890-6820 \AA\ for the 1996-spectra and 4435-6820 \AA\
for the 2000-spectra. The radial velocities listed in the third column
of Table~\ref{vrad} are the averages of the values from the individual
orders; the errors represent the spread around the average. Note that
systematic errors can still be included, e.g. due to the wavelength
calibration. We expect the latter not to exceed 0.75 km s$^{-1}$ (see
Sect.~\ref{highres_obs}). Our measurements confirm that the
narrow-lined component in S\,1082 shows radial-velocity variations of
only a few km s$^{-1}$.

The radial-velocity curves of the stars in the eclipsing binary were
measured after the contribution of the hot, narrow-lined star to the
total spectrum was removed. To that end, all spectra were first
brought to the rest frame of the hot star. Then, a hot-star template
spectrum was constructed by taking the median of the spectra obtained
at phases 0.95, 0.97, 0.0035, 0.019 and 0.048; these particular phases
were chosen in order to let the spectral line profiles of the template
be as symmetric as possible. For the 1996-spectra, taken with a
different instrumental setup, the spectrum obtained at phase 0.092 was
chosen as a template. The relative contribution of the hot-star
template to the remaining spectra was derived with the same fitting
procedure as described in Sect.~\ref{spectra}: the template multiplied
with scaling factors ranging from 0.025 to 1 was subtracted from each
individual spectrum, and the scaling that produces the smallest
residuals around a fit to a straight line was chosen as the
appropriate scaling for that particular spectrum. The scaled template
was then subtracted from the total spectrum to obtain the spectrum of
the binary at various phases.

\begin{figure}
%\resizebox{\hsize}{!}{\includegraphics{plotccf.ps}}
\centering
\includegraphics[width=6cm]{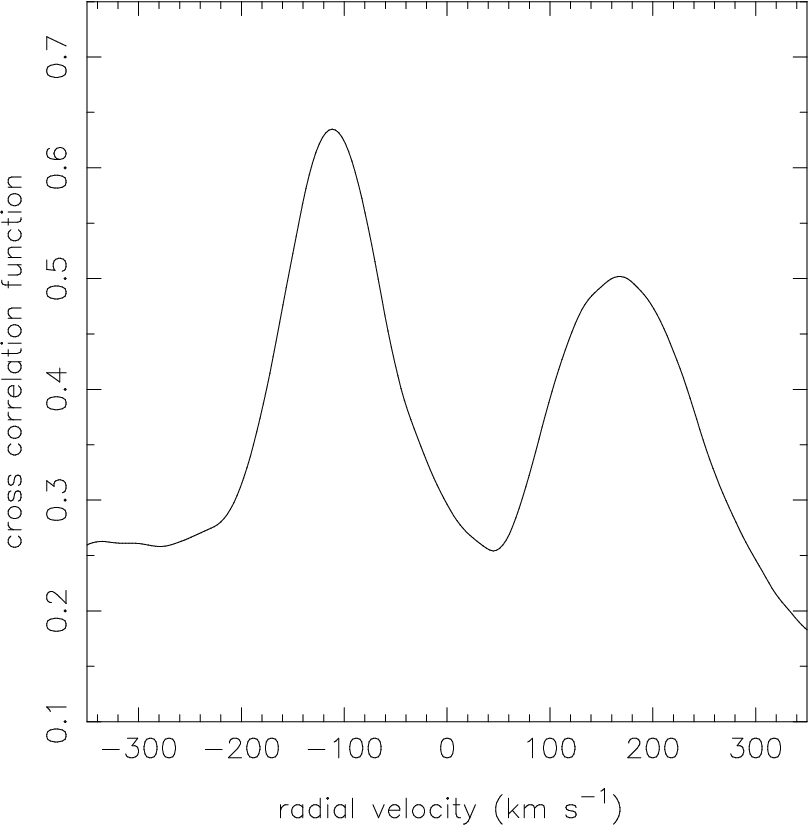}
\caption{Cross correlation function resulting from the residual
spectrum taken at photometric phase 0.2535 after the contribution of
the hot star is removed. The left peak corresponds to the primary star
Aa in the eclipsing binary. The width of peak corresponding to the
secondary Ab is broader which indicates that its spectral lines are
more broadened by rotation. For this particular cross correlation the
spectrum between 5160 and 5240 \AA\ was used. The radial velocity is
given with respect to the hot star B in S\,1082.}
\label{ccf}
\end{figure}

Next, these residual spectra were correlated against a synthetic
spectrum of $T_{\rm eff}$=5250 K and $\log g$=3.5. For this model
spectrum we choose $v_{\rm rot}\sin i$=50 km s$^{-1}$ to roughly match
the apparently broader lines in the spectrum of the binary.  This
procedure was repeated for every order between 4380 and 6435 \AA,
excluding the order that contains the broad H$\beta$ line and the
region between 5690 and 6090 for which the cross correlation functions
are very noisy with no clear peaks. A cross correlation peak is
sometimes visible at $v_{\rm rad}$=0 km s$^{-1}$ and represents
features of the hot star that were not corrected for (or introduced
into the spectrum) by subtraction of the template.  The cross
correlation functions of most spectra clearly show two peaks with a
variable separation (see Fig.~\ref{ccf}). One peak is smaller and
broader than the other, and is redshifted with respect to the template
for phases smaller than 0.5; therefore, this peak is identified with
the secondary star in the eclipsing binary. This is the first
spectroscopic evidence for the third star in S\,1082.  The
measurements of the velocity of the primary near phase 0.1 are
probably distorted due to its eclipse by the secondary star.  In the
following we will refer to the components of the eclipsing binary as
Aa and Ab, and to the outer companion as B.

\begin{table}
\caption{Radial velocities of the three stars in S\,1082, the
components of the eclipsing binary Aa and Ab, and the outer companion
B.  From left to right: heliocentric Julian date (-2\,450\,000) at the
midpoint of observation; photometric phase computed with the ephemeris
of Goranskij et al. (1992); heliocentric radial velocity for star B;
radial velocity of the primary Aa and secondary Ab in the eclipsing
binary with respect to star B in km s$^{-1}$. The spectra of phases
0.5378 and 0.5609 were taken under bad observing conditions.}

\begin{tabular}{rllrr}
HJD-            & phot. & $v_{{\rm rad,B}}$ & \multicolumn{1}{l}{$v_{{\rm rad,Aa}}$} &
\multicolumn{1}{l}{$v_{{\rm rad,Ab}}$} \\
 ~2\,450\,000   & phase & km s$^{-1}$       & km s$^{-1}$ & km s$^{-1}$  \\ 
                &       &                   & & \\
   142.5101 & .09213   &  33.0$\pm$0.1 & -- & -- \\  %r175375
   142.5790 & .1567   &  33.2$\pm$0.1 & -97$\pm$4 & -- \\  %r175384
   142.6837 & .2548   &  32.9$\pm$0.2 & -108$\pm$3 & -- \\  %r175420
  1591.3642 & .9540   &  33.5$\pm$0.4 & -- & -- \\  %r322683
  1591.3822 & .9709   &  33.7$\pm$0.3 & -- & -- \\  %r322684
  1591.4170 & .003451 &  34.2$\pm$0.3 & -- & -- \\  %r322686
  1591.4334 & .01880  &  33.7$\pm$0.3 & -- & -- \\  %r322687
  1591.4640 & .04750  &  32.4$\pm$0.3 & -- & -- \\  %r322691
  1591.4805 & .06294  &  32.8$\pm$0.3 & -- & -- \\  %r322692
  1591.4967 & .07812  &  32.7$\pm$0.3 &  -73$\pm$4 & 122$\pm$7  \\  %r322693
  1591.5452 & .1235   &  33.2$\pm$0.4 &  -96$\pm$3 & 148$\pm$4  \\  %r322699
  1591.5616 & .1389   &  33.3$\pm$0.4 &  -95$\pm$2 & 141$\pm$4  \\  %r322700
  1591.5779 & .1541   &  33.3$\pm$0.4 & -100$\pm$2 & 153$\pm$5  \\  %r322701
  1591.5993 & .1742   &  33.1$\pm$0.5 & -104$\pm$2 & 154$\pm$5  \\  %r322704
  1591.6150 & .1890   &  33.5$\pm$0.4 & -108$\pm$2 & 150$\pm$12 \\  %r322705
  1591.6310 & .2039   &  33.9$\pm$0.5 & -107$\pm$3 & 166$\pm$5  \\  %r322706
  1591.6474 & .2192   &  34.2$\pm$0.5 & -109$\pm$2 & 170$\pm$8  \\  %r322707
  1591.6682 & .2387   &  33.8$\pm$0.5 & -113$\pm$2 & 160$\pm$7  \\  %r322710
  1591.6840 & .2535   &  33.9$\pm$0.5 & -113$\pm$3 & 169$\pm$14 \\  %r322711
  1591.7010 & .2695   &  33.9$\pm$0.5 & -115$\pm$2 & 180$\pm$5  \\  %r322713
  1617.4072 & .3435   &  31.8$\pm$0.4 & -104$\pm$2 & 163$\pm$5  \\  %r326610
  1617.4308 & .3656   &  32.7$\pm$0.5 &  -96$\pm$2 & 153$\pm$5  \\  %r326611
  1617.6147 & .5378   &  36.0$\pm$1.2 &   28$\pm$5 & --         \\  %r326648
  1617.6394 & .5609   &  35.2$\pm$0.8 &   --       & --         \\  %r326659
  1624.3903 & .8832   &  35.4$\pm$0.4 &   91$\pm$3 & -130$\pm$5 \\  %r327495
  1624.4092 & .9009   &  34.6$\pm$0.4 &   85$\pm$4 & -132$\pm$7 \\  %r327499
  1624.5766 & .05765  &  32.4$\pm$0.3 &   --       & --         \\  %r327544
  1624.5949 & .07484  &  32.7$\pm$0.3 &  -80$\pm$4 & 121$\pm$6  \\  %r327545
  1624.6131 & .09188  &  32.6$\pm$0.4 &  -90$\pm$4 & 128$\pm$5  \\  %r327546
\end{tabular}
\label{vrad}
\end{table}

We use the spectrum observed near first quadrature (wavelength region
5160 to 5240 \AA\ containing the \ion{Mg}{I} b triplet) to derive an
estimate for the projected rotational velocities of both components
from the widths of their cross correlation peaks. To that end, we
cross correlate a synthetic spectrum of a certain temperature ($\log
g=4.0$, $v_{\rm rot}\sin i$=30 km s$^{-1}$) with synthetic templates
of the same temperature and surface gravity with $v_{\rm rot}\sin i$
ranging from 10 to 100 km s$^{-1}$, and with the object spectrum. The
peak of the cross correlation functions are fitted with gaussian
profiles. This is repeated for synthetic spectra of temperatures 5000
to 6750 K in steps of 250 K. With the results we construct calibration
curves for each temperature that give the width of the cross
correlation peak as a function of the projected rotational velocity of
the synthetic template. The widths of the cross correlation peaks from
the object spectrum then give a rough estimate for the $v_{\rm
rot}\sin i$ of both stars.  For star Aa we obtain a $v_{\rm rot}\sin
i$ of 51 km s$^{-1}$ (5750 K) to 58 km s$^{-1}$ (6750 K), for star Ab
a $v_{\rm rot}\sin i$ of 83 km s$^{-1}$ (5500, 5750 K) to 87 km
s$^{-1}$ (6750 K).

\begin{figure*}
\centering
      \includegraphics[scale=0.7]{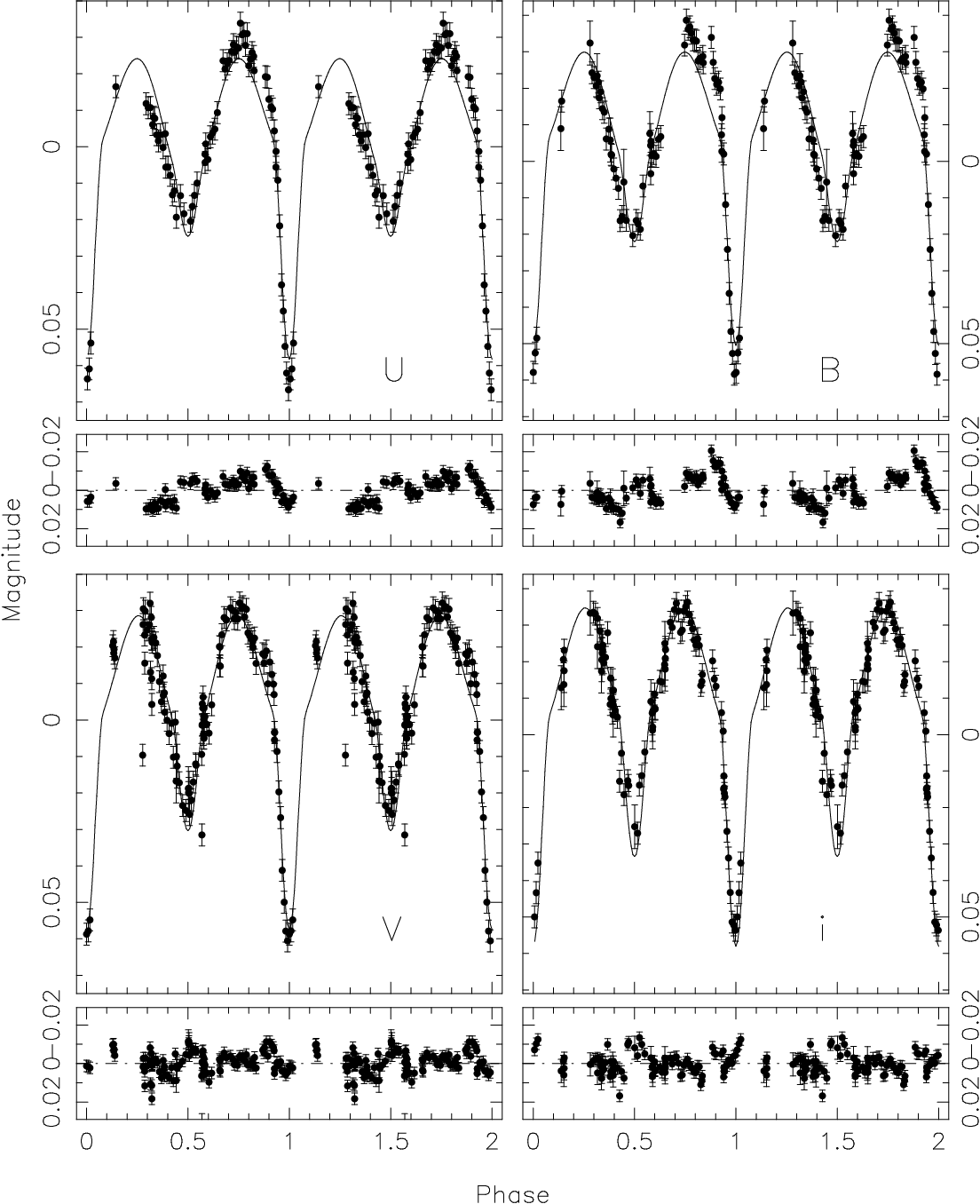}
\caption{Fit to the light curves in the $U$, $B$, $V$ and Gunn $i$ band.
The small panels show the residuals to the fit. Only the data from the
first observation run are used.}
\label{lightcfit}
\end{figure*}

\section{Parameter estimation for the eclipsing binary} \label{lightc}
We used the recently written Eclipsing Light Curve (ELC) code of Orosz
\& Hauschildt (2000) to fit the basic binary observables of S\,1082,
namely the light curves in the four filters, the two velocity curves
and the observed projected rotational velocities of the two stars Aa
and Ab.  The blue straggler B in this case is modelled as ``third
light''.  The addition of a constant (in phase) light will reduce the
observed relative amplitudes of the light curves.  The ELC code uses
model atmosphere specific intensities, so the flux from a third star
can be self-consistently added to several different bandpasses (i.e.\
a blue third light star dilutes the $B$-band light curve of a red
binary more than it dilutes the $I$-band light curve).  We mainly used
the specific intensities derived from the {\sc NextGen} models
(Hauschildt et al.\ 1999a; 1999b).  A few models were also computed
using specific intensities derived from Kurucz (1979) models. Only the
photometric data of run 1 were used for the light curve fitting as
they form the largest set of homogeneous measurements.
\nocite{hausea1} \nocite{hausea2} \nocite{bevi}

%Hauschildt, Peter H.; Allard, France;
%Ferguson, Jason; Baron, E.; Alexander, David R., 1999, ApJ, 525, 871.

%
% Hauschildt, Peter H.; Allard, France; Baron, E., 1999, ApJ, 512, 377
%

We start by estimating the component masses of the eclipsing binary.
Unfortunately, we only have radial velocities near one quadrature
phase (poor weather prevented us from observing the opposite
quadrature), so the velocity amplitudes of the two curves and the
binary systemic velocity will not be as constrained as we would like
them to be.  We fit sinusoids to each of the velocity curves, fixing
the period at the photometric period.
%and forcing the systemic velocity to
%be the same as the mean cluster velocity.  
For star Aa we find $K_{\rm Aa}=120.8\pm 6.4$ km s$^{-1}$,
$\gamma_{\rm rel,Aa}=1.7\pm 5.8$~km s$^{-1}$, $\phi_{\rm
0,Aa}({\rm spect})=0.747\pm 0.007$, where $\gamma_{\rm rel}$ is the systemic
velocity relative to the mean cluster velocity and where
$\phi_{\rm 0}({\rm spect})$ refers to the phase of maximum velocity.  For
star Ab we find $K_{\rm Ab}=190.1\pm 12.5$ km s$^{-1}$, $\gamma_{\rm rel,Ab}=-2.6\pm 12.5$~km s$^{-1}$, $\phi_{\rm0,Ab}({\rm
spect})=0.242\pm 0.010$.  The errors on the individual velocities
were scaled to yield $\chi^2_{\nu}=1$ for each curve, and the error
estimates on the fitted parameters were derived using the scaled
uncertainties.  The phasing of the curves are consistent with
expectations, where star Aa has its maximum velocity one fourth of an
orbital cycle after the deeper photometric eclipse.  Since systemic
velocities are consistent with being zero, we will assume the binary
is a cluster member and hence has $\gamma_{\rm rel} =0$ km s$^{-1}$.
In this case we find for star Aa $K_{\rm Aa} =119.1\pm 3.1$ km
s$^{-1}$ and $\phi_{\rm 0,Aa}({\rm spect})=0.748\pm 0.007$, and for star Ab we
find $K_{\rm Ab}=190.1\pm 11.6$ km s$^{-1}$ and $\phi_{\rm 0,Ab}({\rm
spect})=0.243\pm 0.010$.  Taking the sinusoid fits at face value, we
can immediately compute the mass ratio of the binary.  We find
$Q\equiv M_{\rm Ab}/M_{\rm Aa}=K_{\rm Aa}/K_{\rm Ab}=0.63\pm 0.04$.
The minimum masses of the two component stars are then $M_{\rm
Aa}\sin^3i=2.01\pm 0.38\,M_{\odot}$ and $M_{\rm Ab}\sin^3i=1.26\pm
0.27\,M_{\odot}$ (see Table~\ref{obstab} for a summary of the observed
parameters of the eclipsing binary).

There are a total of twelve free parameters in the model: the filling
factors (by radius) of the two stars $f_{\rm Aa}$, $f_{\rm Ab}$, the
mean temperatures of the two stars $T_{\rm Aa}$, $T_{\rm Ab}$, the
``spin factors'' of the two stars $\Omega_{\rm Aa}$, $\Omega_{\rm
Ab}$, where $\Omega$ is the ratio of the rotational angular velocity
to the orbital angular velocity, the inclination $i$, the mass ratio
$Q$, the orbital separation $a$, the temperature of the third light
star $T_{\rm B}$, the surface gravity of the third light star $\log
g_{B}$, and the third light scaling factor $SA_{\rm B}$.  We assume
the orbit is circular, and that the third light star B does not vary.
The gravity darkening exponent of star Aa was fixed at 0.25, the
standard value for a star with a radiative envelope, while the gravity
darkening exponent for star Ab was set at the standard
convective value of 0.08.  The ELC code uses Wilson's (1990) detailed
reflection scheme, and for this problem the albedo of star Aa was
taken to be 1 and the albedo of star Ab was taken to be 0.5.  Three
iterations of the reflection scheme were needed to achieve
convergence.  Both stars in the binary are assumed to be free of
spots.  A variation of the ``grid search'' routine outlined in
Bevington (1969) was used to optimise the fits.  In practice the
fitting procedure involved a great deal of interaction where some
parameters were temporarily fixed at certain values.  Several
two-dimensional grids in parameter space were defined (for example a
grid of points in the $f_{\rm Aa},f_{\rm Ab}$ plane).  For each point,
we fixed those parameters at the values defined by the grid location
and optimised the other parameters, creating contours of $\chi^2$
values.  New parameter sets were optimised using the set of parameters
for a nearby point that gave the best previous fit.  After the lowest
$\chi^2$ value in a grid was found, then a new grid using other
parameters was computed using the best solution as a starting point.
The fitting took several weeks of CPU time on an Alpha XP 1000, and
sampled a wide range of parameter space.  We are reasonably confident
that our results are at or very near the global $\chi^2$ minimum.

\begin{figure}
%      \resizebox{\hsize}{!}{\includegraphics{plotRV.new.ps}}
      \includegraphics[width=6.5cm]{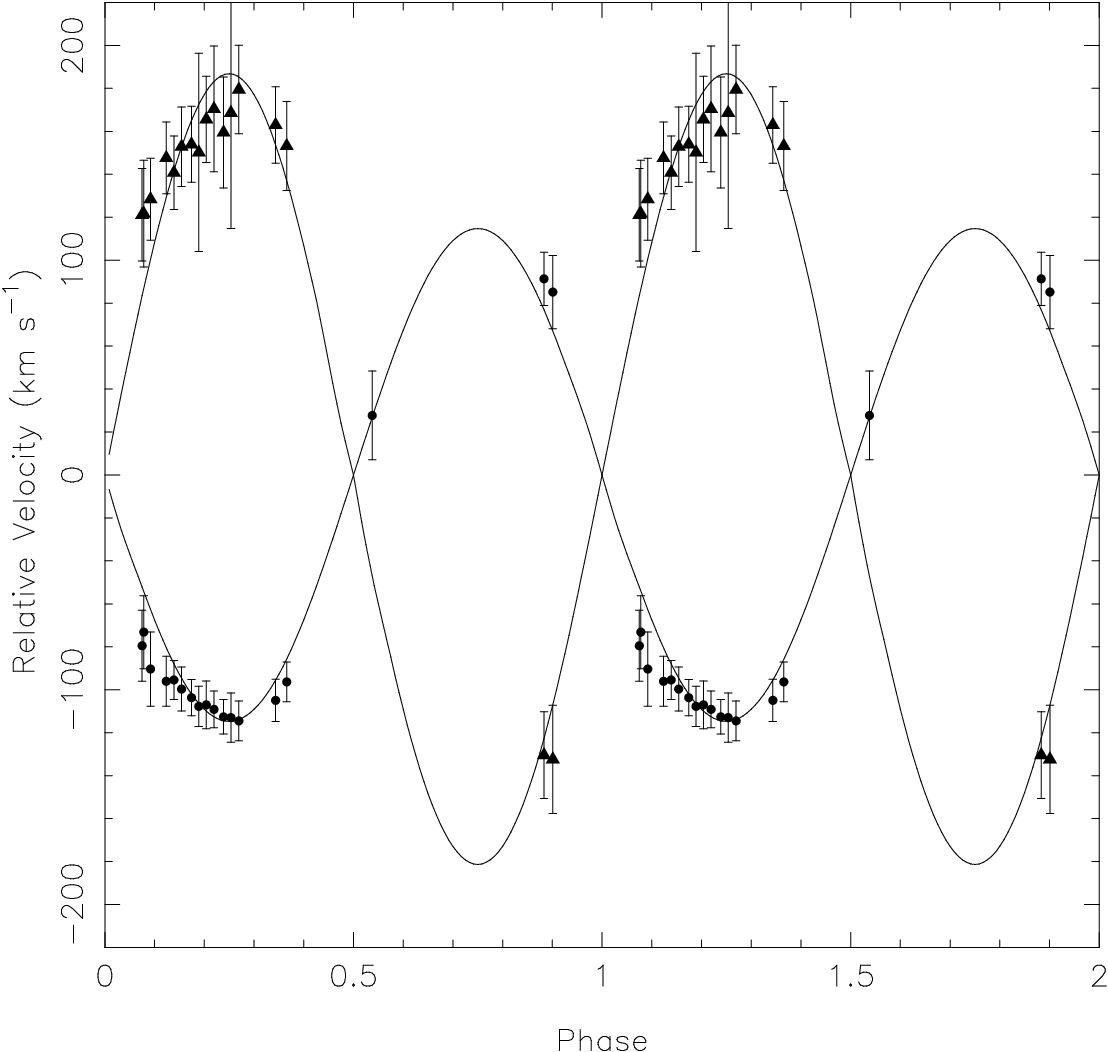}
\caption{Radial-velocity curves of star Aa of the eclipsing binary in
S\,1082 (filled circles) and star Ab (filled triangles). The lines
indicate the best-fitting velocity curves as computed by ELC.}
%to the data with $K_1$=119.1$\pm$3.1 km s$^{-1}$ and
%$K_2$=190.1$\pm$11.6 km s$^{-1}$.}
\label{rvcurves}
\end{figure}

\begin{figure}
%      \resizebox{\hsize}{!}{\includegraphics{plotcomps.new.ps}}
     \includegraphics[width=6.5cm]{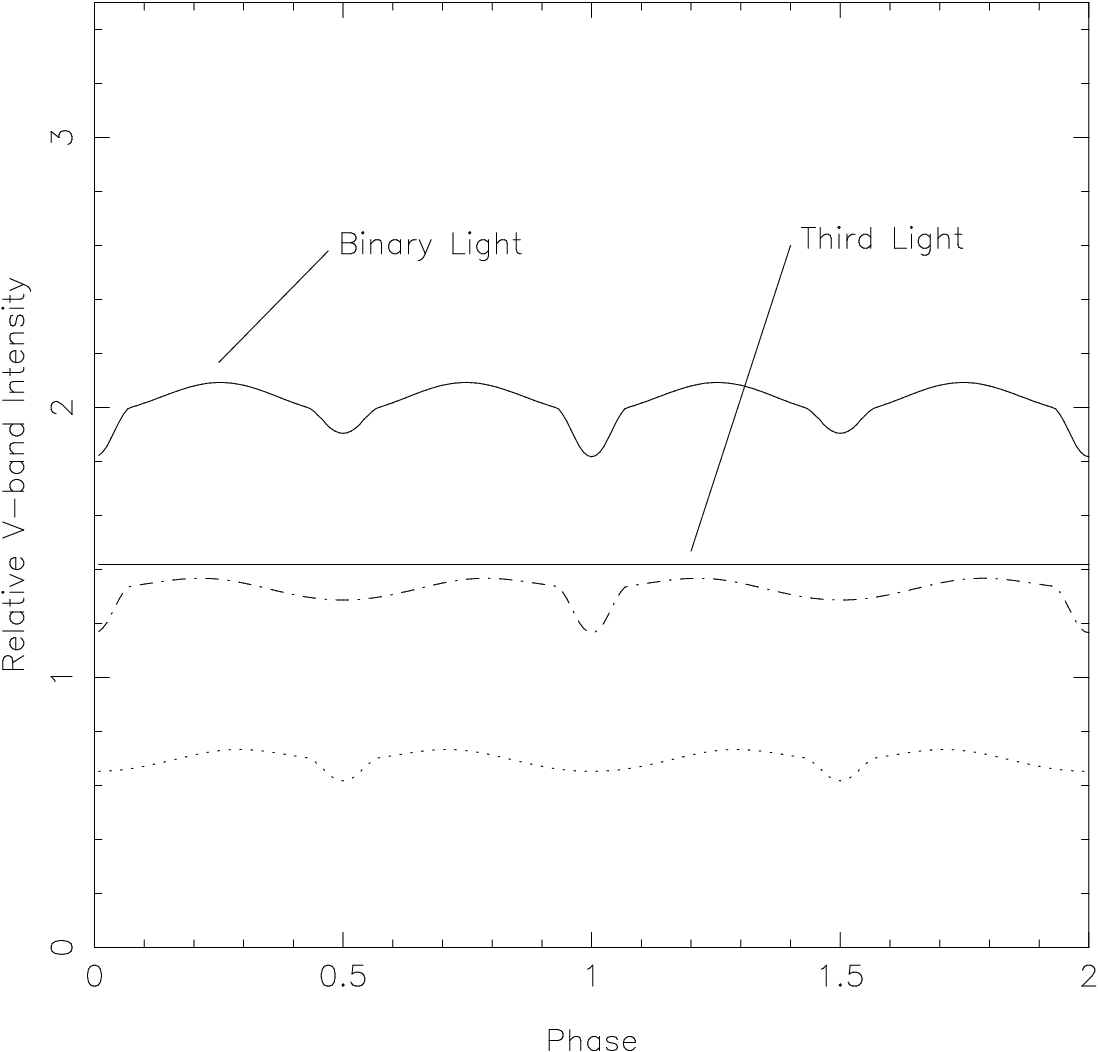}
\caption{Light curves of the three components in S\,1082
(dashed-dotted line indicates the primary Aa in the eclipsing
binary). }
\label{lightcomp}
\end{figure}

\begin{figure}
%      \resizebox{\hsize}{!}{\includegraphics{binary.ps}}
      \includegraphics[width=6.5cm]{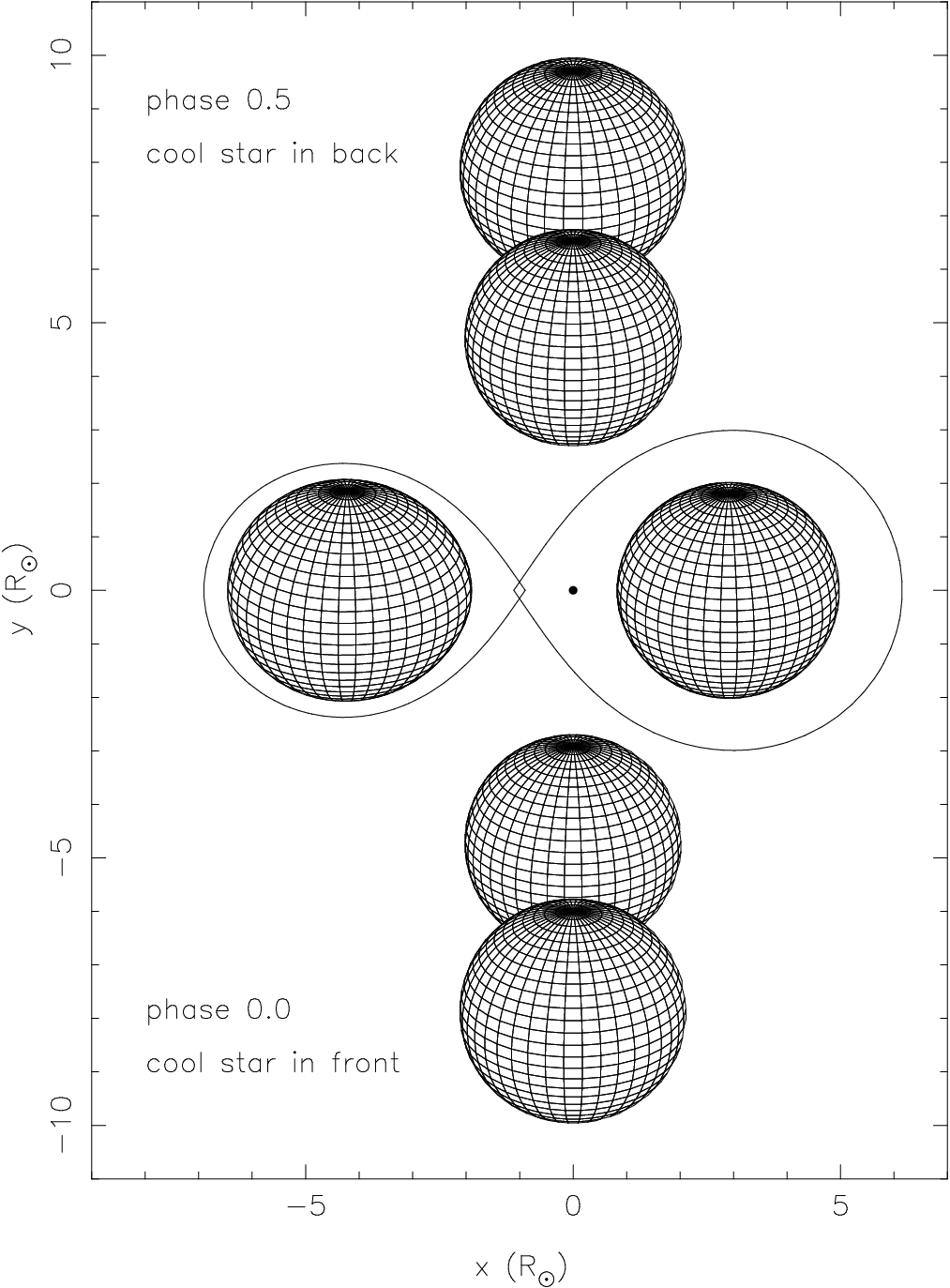}
\caption{Cartoon of the configuration at three orbital phases of the
inner binary according to the model listed in Table~\ref{fittab}.
Note that the Roche-lobes of the two stars Aa and Ab slightly overlap
due to the fact that neither star corotates with the orbit.}
\label{conf}
\end{figure}

We found a relatively large number of solutions with similar $\chi^2$
values.  Fig.~\ref{lightcfit} shows a typical fit.  The light curves
in the 4 bands (Fig.~\ref{lightcfit}) are fitted reasonably well,
although there are still some systematic deviations, especially near
the eclipse phases.  The radial velocities are also fitted reasonably
well (Fig.~\ref{rvcurves}), but again there are some small systematic
deviations (the velocity curves seem to be systematically flatter than
the sinusoid fits near phase 0.25).  In all solutions the binary is
detached, i.e.\ both stars are well within their respective Roche
lobes.  We applied two additional constraints in order to narrow down
the range of parameters.  The first constraint is that the total $V$
magnitude and $B-V$ colour of the model should match the observed $V$
magnitude of S\,1082.  In this case the apparent $V$ magnitude of the
model is easy to compute.  We used the synthetic photometry computed
from the {\sc NextGen} models\footnote{\tt
ftp://calvin.physast.uga.edu/pub/NextGen/Colors/} to compute expected
absolute $V$ magnitude of each star Aa and Ab in the eclipsing binary
from its temperature, radius, and surface gravity.  The $V$ magnitude
of the blue straggler B then follows from the fitted luminosity
scaling.  The second constraint is that the implied mass of the blue
straggler is roughly consistent with its place in the colour-magnitude
diagram.  That is, the radius of the blue straggler B can be computed
from the distance, the $V$ magnitude, and the temperature.  Since the
gravity of the blue straggler B is specified in the models, its mass
can then be computed.  The blue colour of S\,1082 requires a relatively
hot third light star ($\approx 7500$~K), and its surface gravity must
be near $\log g=4.25$ in order for the mass to be near $\approx
1.7\,M_{\odot}$.  The derived astrophysical parameters for the adopted
model are summarised in Table~\ref{fittab}.  The errors on the
parameters were estimated from the $\chi^2$ values generated in the
various grid searches.  These error estimates may be too small, given
the complicated nature of the model.  The errors on the masses were
taken to be on the order of 15\%, as judged from the quality of the
sinusoid fits.  The light curves of the three components are shown in
Fig.~\ref{lightcomp}, and a cartoon of the binary at three phases in
Fig.~\ref{conf}.

%The
%blue straggler contributes about 70\% of the total light in $V$,
%roughly consistent with fraction estimated from spectroscopy.

% All three stars together
%have an absolute $V$ magnitude of about 1.33, and accounting for the
%distance and extinction, the apparent $V$ magnitude would be $\approx
%11.05.$ A 10\% error on the radius of the hotter star would result in
%an error in its absolute magnitude of $\approx 0.2$ mag.  [***these
%will move slightly***] Hence our derived value of $V_{\rm mod}=11.05$
%compares favourably with the observed value of $V_{\rm obs}=11.251$
%(Montgomery et al.\ 1993).
%The total mass of the binary 
%is about $4.4\,M_{\odot}$.  If the
%binary was formed via the straightforward interaction of two single
%stars, then one would expect a total mass smaller than twice the
%turnoff mass of about $2.6\,M_{\odot}$.  Apart from the high mass, the
%model reproduces the observed properties of S\,1082 reasonably well.

We have assumed that the stars are not spotted and that the eclipsing
binary has a circular orbit.  A violation of either one of these
assumptions could alter the light curves to produce the small
systematic deviations seen in the residuals.  Bright or dark spots
could either add or remove light at certain phases, complicating the
analysis.  Given the rather large number of free parameters we have
now we did not consider adding spots at this time.  If this binary is
part of a triple, then the orbit could be eccentric (see
Sect.~\ref{disc}).  A slight eccentricity ($e\approx 0.05$ say) could
cause the maxima to be asymmetric and the minima to be shifted
slightly in phase.  Our current velocity curves do not have enough
phase coverage to place meaningful constraints on the eccentricity, so
any firm conclusions on the eccentricity will have to await the
arrival of additional data.

\begin{table}
\caption{Observed binary parameters.  Aa and Ab refer to the stars in
the eclipsing binary. $V$ and $B-V$ are taken from Montgomery et al.\
(1993).}
\begin{tabular}{lr}
\hline
parameter & value            \\
\hline
$P_{\rm phot}$ & 1.0677971(7) \\
$T_0$ & 2\,444\,643.250(2) \\
$V$ & 11.251  \\
$B-V$ & 0.415 \\
$K_{\rm Aa}$ (km s$^{-1}$) & 119.1(3.1)  \\
$K_{\rm Ab}$ (km s$^{-1}$) & 190.1(11.6)  \\
$Q$ & 0.63(4) \\
$\gamma_{\rm Aa}$ (km s$^{-1}$) & 0.0 [fixed] \\
$\gamma_{\rm Ab}$ (km s$^{-1}$) & 0.0 [fixed] \\
$M_{\rm Aa}\sin^3i$  ($M_{\odot}$) & 2.01(38) \\
$M_{\rm Ab}\sin^3i$  ($M_{\odot}$) & 1.26(27)  \\
$v_{\rm rot,Aa} \sin i$ (km s$^{-1}$) & 56(5)  \\
$v_{\rm rot,Ab} \sin i$ (km s$^{-1}$) & 83(5)  \\
\hline
\end{tabular}
\label{obstab}
\end{table}

%\begin{table}
%\caption{Fitted binary parameters.  Aa and Ab refer to the stars in
%the eclipsing binary, B to the third, outer star.}
%\begin{tabular}{lr}
%\hline
%parameter & value            \\
%\hline
%inclination (deg) & $64.0\pm 1.00$ \\
%%$Q$ & $0.63\pm 0.04$ \\
%$f_{\rm Aa}$   & $0.520\pm 0.010$ \\
%$f_{\rm Ab}$   & $0.700\pm 0.010$ \\
%$R/R_{\rm Roche, Aa}$   & $0.66\pm 0.07$ \\
%$R/R_{\rm Roche, Ab}$   & $0.86\pm 0.07$ \\
%$\Omega_{\rm Aa}$       & $0.49\pm 0.05$        \\
%$\Omega_{\rm Ab}$       & $0.91\pm 0.05$        \\
%temperature star Aa (K) & $6480\pm 25$      \\
%temperature star Ab (K) & $5450\pm 40$      \\
%temperature star B(K) & $7500\pm 50$     \\
%radius star Aa ($R_{\odot}$) & 2.07$\pm 0.07$      \\
%radius star Ab ($R_{\odot}$) & 2.17$\pm 0.03$      \\
%radius star B  ($R_{\odot}$) & $\approx$ 2.5     \\
%gravity star Aa (cgs)    &  $4.21\pm 0.02$ \\
%gravity star Ab (cgs)   &  $4.00\pm 0.02$ \\
%gravity star B  (cgs) & $4.25\pm 0.05$ \\
%mass star Aa ($M_{\odot}$) & $2.70\pm 0.38$ \\
%mass star Ab ($M_{\odot}$) & $1.70\pm 0.27$ \\
%mass star B  ($M_{\odot}$) & $\approx 1.7$ \\
%$V$, $B-V$ star Aa & 12.33$\pm$0.11, 0.51$\pm$0.02 \\
%$V$, $B-V$ star Ab & 13.10$\pm$0.11, 0.82$\pm$0.02 \\
%$V$, $B-V$ star B & 12.24$\pm$0.11, 0.33$\pm$0.01 \\
%$V$, $B-V$ all         & 11.30$\pm$0.11, 0.48$\pm$0.02 \\
%\hline
%\end{tabular}
%\label{fittab}
%\end{table}

\begin{table}
\caption{Fitted binary parameters.  Aa and Ab refer to the stars in
the eclipsing binary, B to the third, outer star.}
\begin{tabular}{lrrr}
\hline
parameter                & Aa         & Ab        & B  \\
\hline
$f$       & 0.520(10) & 0.700(10)  &     \\
$R$ ($R_{\odot}$) & 2.07(7)   & 2.17(3)    & $\approx$ 2.5 \\
%\multicolumn{1}{r}{$R/R_{\rm Roche}$} &  0.66(7) & 0.86(7) & \\
$R/R_{\rm Roche}$ &  0.66(7) & 0.86(7) & \\
$\Omega$                 & 0.49(5)   &  0.91(5)   &        \\
$T$ (K)          & 6480(25)  & 5450(40)   & 7500(50)      \\
log $g$ (cgs)            &  4.21(2)  & 4.0(2)     & 4.25(5) \\
$M$ ($M_{\odot}$)       &  2.70(38) & 1.70(27)   & $\approx 1.7$ \\
$V$                      & 12.33(11) & 13.10(11)  & 12.24(11) \\
$B-V$                    &  0.51(2)  &  0.82(2)   &  0.33(1) \\
\hline
$i_{\rm A}$ (deg) & $64.0(1.0)$ \\
$a_{\rm A}$ ($R_{\odot}$)                     & 7.2(4) \\
$V$, $B-V$ total         & \multicolumn{2}{l}{11.30(11), 0.48(2)} \\
\hline
\end{tabular}
\label{fittab}
\end{table}

\nocite{wils}

\begin{figure*}[t]
      \resizebox{\hsize}{!}{\includegraphics{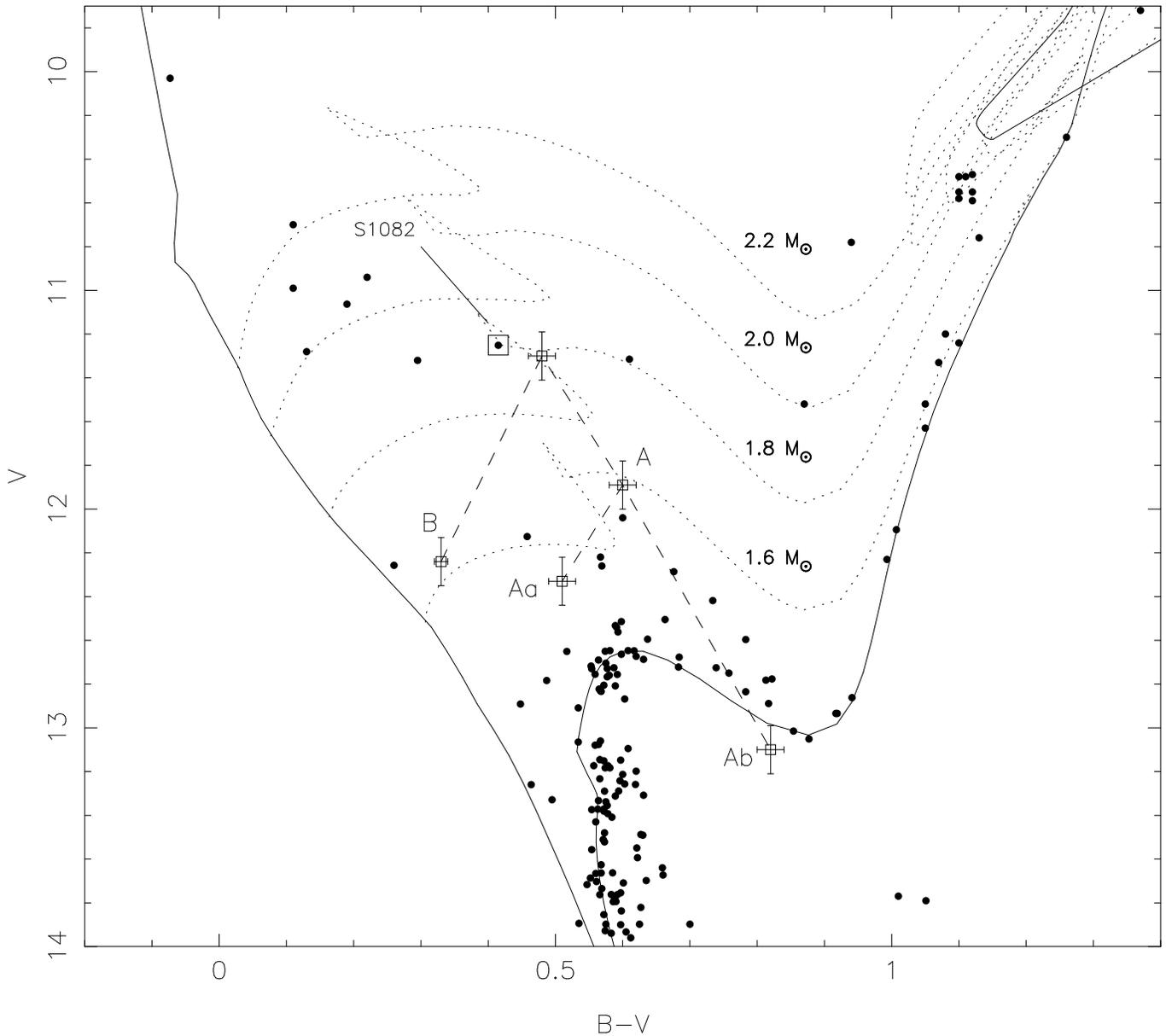}}
\caption{Colour-magnitude diagram of M\,67 that shows the
decomposition of S\,1082 into a blue straggler (B) and stars Aa and Ab
in the eclipsing binary (A); a dashed line connects their
positions. The observed location of S\,1082 is indicated with a box;
the height of the square equals the depth of the primary eclipse. The
dotted lines are evolutionary tracks for 1.6, 1.8, 2.0 and 2.2
M$_{\odot}$ stars corrected for the distance modulus and reddening of
M\,67 ($Z$=0.02, Pols et al. 1998). The 4 Gyr-isochrone is included to
indicate the expected positions of cluster members; the 1 Myr
isochrone is included to give an estimate for the location of the ZAMS
(Pols et al. 1998). $B$ and $V$ magnitudes of M\,67 stars are from
Montgomery et al. (1993). Only stars with a proper-motion membership
probability $>0.8$ (Girard et al. 1989) are plotted.
\label{cmd}}
\end{figure*}

Fig.~\ref{cmd} shows the decomposition of S\,1082 into its components
in a colour-magnitude diagram of M\,67.  Star Ab is located near the 4
Gyr isochrone that provides the best fit to the observed
main-sequence, subgiant and giant stars (Pols et al.\ 1998).  Hence
the position of star Ab is consistent with the expected value based on
its mass.  The positions of star Aa and the blue straggler B are a bit
more uncertain since one can ``trade off'' flux between the two stars
(i.e.\ the blue straggler B can be made brighter at the expense of
star Aa in the binary).  We note that the star Ab is always located
(within the errors) on the 4 Gyr isochrone.  The error bars shown on
the $V$ magnitudes reflect the uncertainties for our adopted model
which produces an overall $V$ magnitude of all three stars close to
the observed value.  In any event, the hotter star Aa is subluminous
by at least 2 mag in V (compare with the isochrone for a 2.2
$M_{\odot}$ star).  The position of the blue straggler lies to the red
of the extension of the main-sequence as defined by the 1 Myr
isochrone.

\nocite{oroshaus} \nocite{gray} \nocite{polsea}

In order to obtain observed times of primary minimum we use the model
light curve as a template to fit the data near the primary minima
observed during run 1 and run 2. Seven additional times of minimum
were obtained from fitting the data in Table~3 of Goranskij et al.\
and from the data points of Simoda (1991) (see first column of
Table~\ref{octab}). A straight line is fitted to the observed times of
minimum to find a new period and $T_0$ (see Table~\ref{obstab}). The
period thus derived is compatible with the period listed by Goranskij
et al.\ (Eq.~\ref{ephem}).  We use the new ephemeris to compute
observed minus computed (O--C) times of primary eclipse and the
corresponding cycle number with respect to $T_0$ (second and third
column of Table~\ref{octab}).  The peak-to-peak amplitude $\Delta$O--C
is $\sim$39 minutes. If these variations are real and caused by the
motion of the eclipsing binary around a third body it would correspond
to a minimum semi-major axis in the outer (o) orbit of the binary (b)
$a_{\rm o,b}\sin i_{\rm o}$=1/2~$c$~$\Delta$O--C=2.3 AU.  Assuming
that the blue straggler (B) has a mass compatible with its position
in the colour-magnitude diagram (about 1.7 M$_{\odot}$) this
corresponds to a minimum value of the semi-major axis of the total
system $a_{\rm o} \sin i_{\rm o}$=$a_{\rm o,b}\sin i_{\rm
o}$~(1+$M_{\rm A}/M_{\rm B}) \approx$8.4 AU; combined with the total
mass of the system of 6.1 $M_{\odot}$ and Kepler's third law this
gives a minimum period of 10 years. This is not compatible with the
period of $\sim$3 years found by Milone (1991). We conclude that not
all the variation in O--C is due to light-travel time in the outer
orbit. There is no evidence for periodicity in the O--C times,
although the time baseline is somewhat short and the coverage is
somewhat spotty.

\nocite{simo}

\begin{table}
\caption{Times of primary minimum. From left to right: observed time
of minimum; O--C difference (in days) between the observed and computed
time of eclipse; cycle number of the eclipse with respect to the $T_0$
of Table~\ref{obstab}. The errors in the observed times of minimum and
the O-C values are 0.005 days.}
\begin{tabular}{rrrl}
HJD &  O-C & cycle & source\\
-2\,440\,000 & & & \\
\\
3191.036 & -0.010 & -1360 & Simoda (1991)$^{a}$ \\
4643.250       &  0.000 & 0     & Goranskij et al.\ (1992)  \\
5325.586       &  0.013 & 639   & idem              \\
6773.492       & -0.013 & 1995  & idem              \\
7861.605       &  0.014 & 3014  & idem              \\
7920.333       &  0.014 & 3069  & idem              \\
7944.869       & -0.010 & 3092  & idem$^{b}$              \\
11218.744      & -0.001 & 6158  & run 1             \\
11539.078      & -0.006 & 6458  & run 2             \\
\\
\end{tabular}

$^{a}$Simoda observed two consecutive primary eclipses. These data
were combined to measure one time of minimum. \\ $^{b}$Based on a
measurement of a secondary eclipse which we convert to a primary
eclipse by adding half a period. We note that if the orbit of the
eclipsing binary is eccentric, the two eclipses need not be separated
by half a period.
\label{octab}
\end{table}

\section{Discussion}\label{disc}

Our new observations of S\,1082 confirm the eclipses reported by
Goranskij et al. (1992), the small radial-velocity variations of the
narrow lines in the spectrum (Mathieu et al.\ 1986) and the
time-variation of a broad-lined component seen in high-resolution
spectra (van den Berg et al.\ 1999, Shetrone \&\ Sandquist 2000). In
addition, we now have found that the radial-velocity shifts of the
broad-lined component {\em and} of a third component in the spectra
vary on the photometric period. This clearly demonstrates that the
narrow-lined star, which is a blue straggler on it own, is not part of
the eclipsing binary -- this solves the seeming contradiction in the
properties of this system.  The broad variable component of the
H$\alpha$ line discussed by van den Berg et al.\ (1999) is the
Stark-broadened H$\alpha$ line of the hot component Aa of the inner
binary. The ultraviolet flux measured by Landsman et al. (1998),
higher than expected for a star at the $B-V$ colour of S\,1082, is
only partly explained as a consequence of the bluer colour, i.e.\
higher temperature of the third star.

Comparison with Figs.~2 and 3 of Dempsey et al.\ (1993) shows that the
X-ray luminosity of S\,1082 (Belloni et al.\ 1998) is typical for a
subgiant in an RS\,CVn system. The X-ray emission could therefore be
caused by magnetic activity in the rapidly rotating subgiant Ab. Thus,
our model can explain the X-rays of S\,1082.

Our solution to the photometry is symmetric (see Fig.~\ref{lightcomp})
and does not explain the asymmetry between phases 0.25 and 0.75; nor
the variability in the form of the secondary eclipse between our
first two observations runs. This indicates that variable spots are
present and accordingly that the best-fit parameters of the system are
subject to some additional uncertainty. However, we do not think our
main conclusions are affected.

\nocite{landea98} \nocite{bellea} \nocite{dempea} \nocite{sandshet}

We think that the eclipsing binary forms a bound triple with the third
star, for two reasons. First, the radial-velocity measurements of the
third star B -- the narrow-lined system in S\,1082 -- indicate that it
is in a $\sim$1000\,day orbit around a companion. The systemic
velocities of both the 1000\,day and the 1.07\,day binaries are
compatible with the radial velocity of the cluster. A chance alignment
of two such bright cluster members is unlikely. Second, if we relax
the constraint that the binary is at the cluster distance in solving
the photometry, we find that the best solutions tend to be an Algol
binary at higher mass ($\gtrsim 4\,M_{\odot}$) and larger distance
(about twice the cluster distance). A high-mass binary at such a large
distance from the galactic disk is unlikely.

Assuming then that the binary forms a hierarchical triple with the
third star, in M\,67, we note that the 1-$\sigma$ lower bound to the
mass of the binary is at about three times the turnoff mass of the
cluster. The formation of the binary must thus have involved at least
three stars. The third star is a blue straggler on its own account,
and thus according to most current models its formation involves two
stars. We have to conclude that the formation of S\,1082 required the
interaction of no less than five stars!

This interaction may have started with a binary-binary encounter, in
which two stars collided directly and merged; one of the remaining two
stars ended in a close orbit around the merger, the fourth star in a
wide orbit around the inner binary. The fourth star must be a blue
straggler, which either was already present in the original encounter
or was later exchanged into the system.  That triples are formed
easily by binary-binary encounters, and that they live long enough to
undergo subsequent exchange encounters with a binary, is shown for
example by the computations of Aarseth \&\ Mardling (2001).  It is
less obvious that mergers are common in such encounters, as most
binary-binary encounters are between relatively wide systems.  Merger
products tend to be subluminous (Sills et al.\ 2001), in accordance
with the properties of the more massive star Aa in the inner binary.
The main problem with this scenario for the formation of the S\,1082
system is that a merger product is only subluminous for a very limited
period; comparable to its thermal time scale. This reduces the
available time within which the outer star of the first encounter is
exchanged with a binary or blue straggler in a subsequent encounter,
and indicates that the blue straggler was present in the initial
binary-binary encounter. Alternatively, an initial encounter between
two triple systems can have led immediately to the currently observed
configuration.  In any case, the probability of catching the merger
product while it is strongly subluminous is disconcertingly small.
\nocite{aarsmard} \nocite{sillea}

Aarseth \&\ Mardling (2001) note that the inclination of the outer
orbit in a hierarchical triple with respect to the inner orbit can
induce a large eccentricity in the inner orbit; subsequently, tidal
forces in the inner orbit cause it to shrink.  Thus, the inner orbit
may be eccentric, and smaller than in the past. Our solution for the
inner binary implies that the cool star has a radius which is a
sizeable fraction of the binary separation $R_{\rm Ab}/a_{\rm A} \approx 0.3$.
According to Eq.~2 of Verbunt \&\ Phinney (1995) the circularisation
time scale of the inner binary A is $\sim$ 10$^3$ year; the
eccentricity of A and asynchronicity of Ab thus will be determined by
the competing effects of the perturbation by the outer star B and the
tidal forces in the inner binary. Only if the latter are substantially
reduced with respect to the Zahn (1977) formulation -- as suggested by
e.g. Goodman \&\ Oh (1997) -- do we expect measurable eccentricity and
asynchronicity.

\nocite{verbphin} \nocite{zahn77} \nocite{goodoh}

Whereas our new observations have allowed us to resolve the apparent
contradictions in the earlier data and to determine the system
parameters, we conclude that these parameters -- in particular those
of the primary Aa of the inner binary -- are hard to understand in
terms of standard stellar and binary evolution, even when stellar
encounters and mergers are taken into account. This situation is
remarkably similar to that in the study of two other members of M\,67,
located below the subgiant branch in the colour-magnitude diagram,
both of which turn out to be binaries (Mathieu et al.\ 2001, in
preparation); and indeed to our lack of understanding of the blue
straggler population in the cluster.  Many of these stars may be
mergers, which raises the question whether the object that results
when two stars merge may follow an evolutionary track which is very
different from that of an ordinary star of the same mass; and whether
they can do so for a period of time which significantly exceeds the
thermal time scale.

Further studies of the remarkable triple system S\,1082 should provide
a better-sampled radial-velocity curve, necessary to determine whether
the inner binary is eccentric, and to measure its systemic velocity
more accurately to establish cluster membership. Better-quality light
curves may also serve to improve the accuracy of the inner orbit,
e.g. if the inner orbit is eccentric the separation of the eclipses is
different of 0.5. A long-term sampling of the light curves will be
required to improve the interpretation of the O--C in the timings of
the primary eclipse.

\begin{figure}
\resizebox{\hsize}{!}{\includegraphics{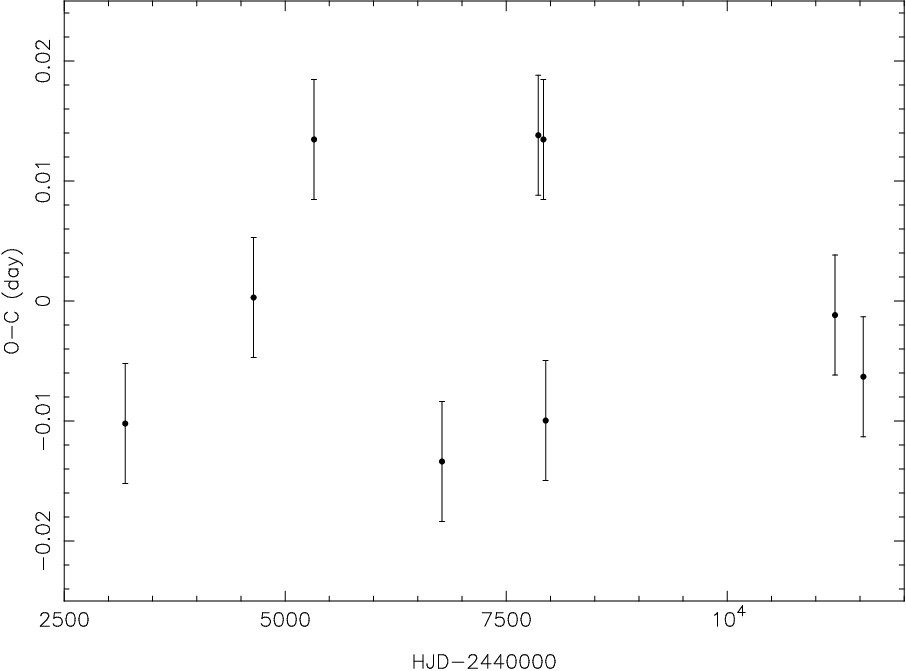}}
\caption{O--C times of the primary eclipse in days versus heliocentric
Julian date (-2\,440\,000) of the measurement.}
\end{figure}

\begin{acknowledgements}
The authors wish to thank Magiel Janson, Rien Dijkstra, Gertie
Geertsema, Remon Cornelisse and Gijs Nelemans for doing part of the
observations. We also want to thank David Latham for discussions. MvdB
is supported by the Netherlands Organization for Scientific Research
(NWO).
\end{acknowledgements}

%\bibliographystyle{apj}
%\bibliography{biball}

%\listofobjects

\end{document}